**Investigating biomechanical determinants of endothelial permeability in a modified hollow fibre bioreactor**

**Stephen G. Gray[1], Peter D. Weinberg***

Department of Bioengineering
Imperial College London
London SW7 2AZ, UK

Short title: Biomechanics of endothelial permeability

*Corresponding author
Professor P. D. Weinberg
Department of Bioengineering
Imperial College London
London SW7 2AZ
United Kingdom
p.weinberg@imperial.ac.uk

[1]Current address
NuoBio
1810 Malcolm Ave #204,
Los Angeles
California 90025
USA

# Abstract

Effects of mechanical stress on the permeability of vascular endothelium are important to normal physiology and may be critical in the development of atherosclerosis, where they can account for the patchy arterial distribution of the disease. Such properties are frequently investigated *in vitro*. Here we evaluate and use the hollow fibre bioreactor for this purpose; in this system, endothelial cells form a confluent monolayer lining numerous plastic capillaries with porous walls, contained in a cartridge. The capillaries were perfused with a near-aortic waveform by an external pump, and permeability was assessed by the movement of rhodamine-labelled albumin from the intracapillary space to the extracapillary space. Confluence and quiescence of the cells was confirmed by electron microscopy and measurements of glucose consumption and permeability. The system was able to detect previously established influences on permeability: tracer transport was increased by acute application of shear stress and decreased by chronic shear stress compared to a static control, and was increased by thrombin or an NO synthase inhibitor under chronic shear. Increasing viscosity by addition of xanthan gum reduced permeability under both acute and chronic shear. Addition of damping chambers to reduce flow pulsatility increased permeability. Modifying the cartridge to allow chronic convection across the monolayer increased effective permeability more than could be explained the addition of convective transport alone, indicating that it caused an increase in permeability. The off-the-shelf hollow fibre bioreactor provides an excellent system for investigating the biomechanics of endothelial permeability and its potential is increased by simple modifications.

# 1. Introduction

Transport of macromolecules across vascular endothelium is essential for normal physiological functions such as the delivery of lipids, metal ions, hormones and other signalling molecules to cells. Abnormal transport appears to be important in pathological processes. A case of particular interest is atherosclerosis, which is triggered by elevated transendothelial transport of low density lipoprotein. The patchy distribution of atherosclerosis within the arterial tree has been attributed to influences of spatially varying mechanical stresses on endothelial permeability [1-4].

Assessment of unmodified mechanical stresses and permeability *in vivo* can only determine correlations between the two. Establishing causality and mechanisms requires intervention. Previous work has modified mechanical stress – principally haemodynamic wall shear stress – by ligating vessels *in vivo* [5,6] or by artificial perfusion of vessels *in situ* or *ex vivo* [7,8]. Fewer such studies have been conducted that their potential pathological importance would merit, reflecting the substantial technical difficulties. Many studies have instead investigated the link by applying stresses to endothelial monolayers *in vitro*.

Methods for applying fluid shear stresses *in vitro* while measuring permeability include culturing cells on porous microcarrier beads in a column [9], on a porous membrane under a rotating disc [10], and in wells swirled on an orbital shaker or in microchannels, both with tracers that bind to a substrate [11, 12]. Effects of pressure-driven transendothelial flow rather than luminal flow have been studied using porous membranes in custom apparatus [13]. Here we evaluate and use the hollow fibre bioreactor.

The apparatus, introduced by Knazek et al. [14], consists of multiple plastic capillary tubes ("hollow fibres") with porous walls encased in a cartridge. Manifolds at each end of the cartridge allow perfusion of the capillaries via an external circuit that includes a pump and oxygenator; the whole apparatus is maintained in a cell culture incubator. Such devices are now available commercially. Growing cells in the extracapillary space of the cartridge mimics tissue architecture and allows the collection of cell-secreted products. The high surface area-to-volume ratio and porosity of the capillary walls gives excellent solute transport [15]. Alternatively, endothelial cells [16] and epithelial cells [17] can be grown on the *inside* of the capillaries, forming confluent monolayers on the luminal surface of the wall. Introducing tracer into the intracapillary space and monitoring its emergence into the extracapillary space allows the measurement of permeability in monolayers exposed to shear.

The system has been used to determine effects on permeability of co-culture with other cell types [16,18,19] or of conditioned medium [19 again], to determine whether transport is receptor mediated or not [20], and to assess transendothelial movement of cells [21] as well as solutes. The system has also been used to examine influences of different shear levels on biological properties such as protein secretion [22] or proliferation [23], but we are not aware of any previous studies that employed hollow fibre bioreactors to examine effects of different mechanical stresses on endothelial permeability to solutes.

## 2. Methods

*2.1 Isolation and Characterisation of Porcine Aortic Endothelial Cells*

Descending thoracic aortas of Landrace Cross pigs aged 4-6 months were obtained at abattoir and stored in Hanks Balanced Salt Solution (HBSS) containing penicillin (200 U/ml), streptomycin (200 µg/ml), amphotericin (5 µg/ml), and gentamycin (100 µg/ml) for up to 24 h. Porcine aortic endothelial cells (PAECs) were isolated by perfusing the lumen with collagenase as previously described [24]. They were then resuspended in Dulbecco's Modified Eagle Medium (DMEM, Sigma-Aldrich) supplemented with 20% Fetal Bovine Serum (FBS, Sigma-Aldrich), penicillin (100 U/ml), streptomycin (100 µg/ml), amphotericin (2.5 µg/ml), gentamycin (50 µg/ml), 5mM L-glutamine, and 5 µg/ml endothelial cell growth factor (ECGF, Sigma-Aldrich). This solution is referred to in later sections as DMEM with supplements. Cells were cultured under 5% $CO_2$ at 37 C. Medium was changed after 1 h to reduce smooth muscle cell contamination, after 24 h and then every 2 days. Cells were used at passage 1-3.

Endothelial cell purity was assessed from the internalization of DiI-labelled acetylated LDL (Molecular Probes), which was added to confluent monolayers at a final concentration of 10 µg/mL and incubated at 37 C for 12 h. After fixation in 4% paraformaldehyde and nuclear staining with DRAQ5 (BioStatus, 1:1000), the cells were imaged using a laser scanning confocal microscope (Leica SP5) with a 10x 0.40NA objective. DiI-acetylated-LDL was excited at 514 nm and detected between 539-593 nm; equivalent wavelengths for DRAQ5 were 633 and 675-725 nm. The fraction of cells taking up DiI-acetylated-LDL was determined with a manual counting program developed in MATLAB (The MathWorks, Inc).

*2.2 Bioreactor Characteristics, Activation and Coating*

The hollow fibre bioreactor cartridges (FiberCell®) contained 20 polysulphone plus capillaries, each around 10 cm long with 700 µm inner diameter, wall thickness of 300 µm and pore size 0.1 µm. Total surface area was 70 $cm^2$, supporting approximately $10^8$ cells. Flow was provided by a reservoir of medium connected to positive displacement pump in a closed circuit, and gas exchange was enabled by a coil of gas-permeable tubing (FiberCell®) – see Figure 1.

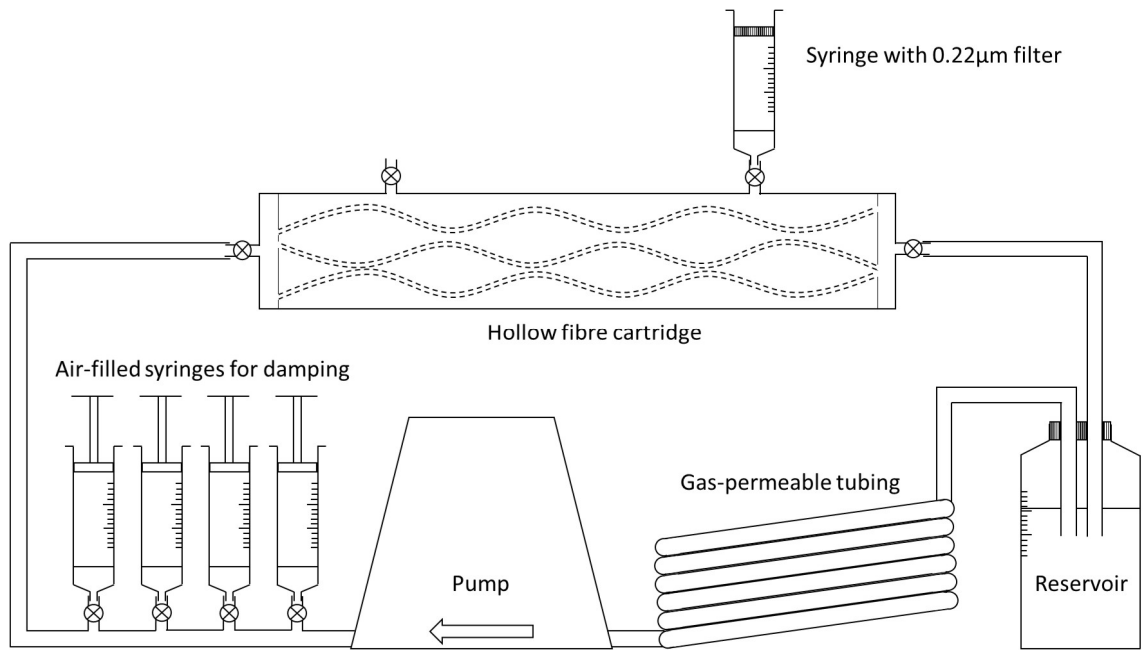

**Figure 1. Diagram of the hollow fibre bioreactor configuration**

Air-filled syringes for damping and the syringe with the 0.22 m filter are modifications of the commercial apparatus and were used only in experiments assessing permeability under steady flow and transendothelial flow, respectively. For experiments under static conditions, fluid was circulated in the extracapillary space rather than the intracapillary space, by connecting the tubing to the two ports on the top of the cartridge. For clarity, only 3 of 20 capillary fibres are shown within the cartridge.

Cartridges, connectors, and the reservoir were sterilized for approximately 2 h with UV light prior to cell seeding. In a laminar flow hood, the entry ports were opened and the fibres perfused with 70% ethanol for 1 min and 5% gelatin for 10 mins. The cartridge was rotated 180° and left for a further 10 minutes before being placed in the incubator for 1 h. The cartridge was rotated 180° every 15 minutes in the incubator. The hollow fibres were then flushed with PBS and the reservoir, filled with 200ml of DMEM with supplements, was attached to the bioreactor circuit, before the system was in turn attached to the pump (FiberCell). DMEM with supplements was pumped through the gas permeable tubing at a flow rate of 15ml/min for 2 h prior to cell seeding.

*2.3 Seeding the Bioreactor*

A confluent T75 flask of PAECs was trypsinized and the solution centrifuged at 220 x *g* for 5 minutes. The pellet was resuspended in 5 ml of DMEM with supplements and the cells counted (normally 5-8 x $10^6$ cells). In a laminar flow hood, the cell suspension was slowly introduced into the intrafibrillar space, whilst the extra volume of fluid this introduced was allowed to permeate the fibre walls and leave from the extrafibrillar space.

The bioreactor was returned to the incubator for 1 h during which time the cartridge was rotated 180° every 10 minutes. The bioreactor was then removed from the incubator and fresh medium slowly flushed through the intrafibrillar space. Loose cells collected in the flushing solution were counted to determine the seeding efficiency (normally 40-60%). The seeding procedure was then repeated using a second T75 flask. In this step, which reduced the time required to achieve confluence, the seeding efficiency was approximately 40-50%.

The pump and gas permeable tubing were attached to the extrafibrillar circuit, the flow rate set to 5 ml/min, and the bioreactor returned to the incubator for 48 h, with medium in the intrafibrillar and extrafibrillar circuits being changed every 24 h. The pump and gas permeable tubing were then attached to the intrafibrillar circuit; the flow rate of 5 ml/min was maintained for 24 h and then increased to 15 ml/min. The peristaltic pump in combination with its system of one-way valves gave a pulsatile flow waveform.

*2.4 Assessing Endothelial Cell Confluence*

Glucose concentrations were monitored to determine when the monolayer had reached confluence since glucose consumption is reduced on contact inhibition and quiescence [25]. Samples of medium were taken from the bioreactor exposed to flow during the first 10-14 days after seeding. Glucose was determined using glucose detection strips and a GluCell® meter.

The permeability of the cells to rhodamine-labelled albumin was measured soon after seeding and at several times after glucose measurements indicated a quiescent monolayer, to determine whether permeability had reached a steady, low level.

To image the monolayers after permeability experiments had been completed, PAECs were fixed for 1 h with 15% paraformaldehyde, introduced by using careful syringe movements from port A to port D, and then washed three times with PBS. The bioreactor was opened with a pipe cutter; individual fibres were then removed from the cartridge and sectioned longitudinally with a razor blade and custom jig. Sectioned fibres were dehydrated in ascending concentrations of ethanol. The ethanol was replaced with $CO_2$ by critical point drying. Samples were then placed on conducting carbon pads on aluminium stubs, coated with gold and imaged by scanning electron microscopy (SEM) using a JEOL JSM 5610 LV with an accelerating voltage of 20 kV and emission current of 10 µA.

*2.5 Preparation of Tracer*

Detailed methods are given elsewhere [26]. Briefly, bovine serum albumin was mixed with sulphorhodamine B acid chloride and then purified of free dye and suspended in a 1:10 dilution of Tyrode's Salt Solution in water by gel filtration, lyophilized, and stored at -20 C. Before use, the conjugate was reconstituted in water to 1/10th of its original volume, further purified with neutralised activated charcoal (0.35 g/g of protein), and sterile filtered.

*2.6 Measuring Monolayer Permeability*

A permeability measurement was performed on the day following seeding to obtain a baseline reading. Further measurements were made once the glucose concentration had decreased from an initial 100 mg/dl to 40-50mg/dL under continuous flow. Any permeability values that were similar to the permeability observed on the first day after seeding prompted an increase in cell culture time and if no change was observed then the experiment was discarded.

The serum content of the culture medium in the reservoir was reduced from 20% to 10% following 10 days of culture and reduced to 4% 24 h before each permeability experiment, in both the intracapillary and extracapillary compartments. For the tracer experiment itself, the solution in the reservoir and gas permeable tubing was replaced with DMEM, supplements other than FBS, 4% BSA, and 1 mg/ml rhodamine-labelled albumin. The tracer solution was circulated for approximately 1 h in all experiments with flow. For static studies, tracer was not circulated but simply added to the intracapillary space using syringes and left there for 1 h. Samples were taken from intracapillary and extracapillary compartments and tracer fluorescence in them was measured using a fluorimeter (model 6285, Jenway) with excitation and emission wavelengths of 570 and 600nm, respectively. The concentration of rhodamine albumin in each diluted sample was determined from a standard curve.

Permeability (P) is defined as the flux of solute (Js) across a known area of membrane (S) per unit concentration difference across the membrane ($\Delta C$):

$$P = J_s / (S \times \Delta C)$$

The term is generally used to refer to diffusive transport, but a vesicular component cannot be ruled out in these experiments (see below). Js was determined from the concentration of rhodamine albumin in the extracapillary space (CECS), the volume of the extracapillary space (VECS), and the time after the addition of the tracer to the intracapillary space (t):

$$J_s = (C_{ECS} \times V_{ECS}) / t$$

The low concentration of tracer in the extracapillary space was ignored when calculating the concentration gradient.

The resistance of the membrane without cells and of the membrane with the cell monolayer, obtained for the same bioreactor, were calculated as reciprocals of permeability. (The average permeability value for the hollow fibre membrane alone was $5.05 \times 10^{-06}$ cm/s.) The former was subtracted from the latter to give the resistance of the cell monolayer alone, and its reciprocal gave the permeability of the monolayer.

Several permeability measurements under different conditions were conducted in each bioreactor. Experiments ended 12-14 days after the first permeability experiment.

*2.7 Effect of Thrombin and L-NAME on Endothelial Permeability*

In the first series of experiments, agents known to affect endothelial permeability were added to the system to determine whether the expected endothelial response could be detected. An absence of effect might indicate that the permeability was dominated by the resistance of the fibre wall, for example as a result of a failure to form a confluent monolayer with tight junctions. The experiments were carried out continuous pulsatile flow, termed "chronic shear stress" (CSS; see next section). Thrombin (Sigma-Aldrich; 10 U/ml) was added for a period of 1 h before permeability was measured. Production of NO was inhibited by the addition of the L-arginine analogue $N^{\omega}$-nitro-l-arginine methyl ester (L-NAME; Sigma-Aldrich; 500 µM) for 24 h prior to the permeability measurement.

*2.8 Permeability under Chronic Shear Stress, Acute Shear Stress and Static Conditions*

In the second series of experiments, the PAECs within the bioreactor were exposed to different flows or no flow, again to determine whether established responses observed in other systems could be replicated. Following the initial permeability measurement under CSS, a further measurement was made after exposure to static conditions for 2-4 days, then after a short period of acute shear stress (ASS) and finally after a repeat of the initial CSS condition. This procedure was repeated 2-3 times per bioreactor.

For CSS and ASS, medium was pumped through the capillaries at an average total rate of 15 ml/min. This gives a wall shear stress (WSS) of 3.75 dynes/cm$^2$ according to the manufacturer. The flow rate was confirmed by use of a transit-time ultrasound flow probe (Transonics) coupled to an analogue-to-digital converter and data analysis system (Notochord). Flow was maintained for 3-10 days at a time for CSS, depending on the interchanges between experimental conditions, and for 4 h after static conditions for ASS.

For static conditions, syringes were filled with DMEM and supplements and attached to either end of the bioreactor cartridge to ensure that cells had sufficient medium. The pump and gas permeable tubing were used to perfuse the extracapillary space. Thus cells on the fibres acquired nutrients and oxygen, and had waste products removed, but the cells were shielded by the fibre wall from the direct influence of flow. This method has been used by Milanova et al in their study of effects of shear on endothelial proliferation [23]; they demonstrated there was no change in oxygen delivery to the endothelial cells compared to intracapillary perfusion.

Flow was maintained at 5 ml/min for approximately 3-5 days prior to permeability measurement. Note that mixing effects of the flow will have had a negligible influence on transport because the concentration of tracer was always much lower in the extrafibrillar than the intrafibrillar space.

*2.9 Permeability with Increased Viscosity*

Experiments were conducted in new bioreactors using medium that had been modified to obtain rheological properties analogous to those of blood. Following the technique outlined by van den Broek et al. [27], xanthan gum (Sigma-Aldrich, 0.69 g/L) was UV sterilized, added slowly to DMEM plus supplements and stirred for 24 h to create a medium with viscosity of 4 x 10$^{-3}$ Pa.s. The sequence of static, ASS and CSS experiments was then repeated in several bioreactors using this medium.

*2.10 Permeability in Modified Bioreactors*

A third series of experiments compared the normal pulsatile flow (CSS) with chronic flow in which almost all the pulsatility had been damped ("steady flow"), and also compared it with CSS plus transmural flow. The experiments required modification of the bioreactor. In order to damp the pulsatility of the flow, four 50 ml syringes filled with air were connected to the flow circuit (Figure 1). Damping was confirmed with the transit-time flow meter.

For the transmural flow experiments, one of the ports accessing the extracapillary space was opened and a syringe with a 0.22-µm filter replacing the plunger was added to the port to expose the system to atmospheric pressure (Figure 1). This modification created a pressure difference across the capillary wall and permitted transmural flow whilst maintaining sterility. After several hours, the transmural flow across the capillary walls and into the cartridge and syringe stabilized to approximately 0.1 mL/h. The syringe was replaced every few hours as the rising level of medium in it would otherwise have significantly altered the transmural pressure difference. All fluid entering the extrafibrillar space and the attached syringe was collected, its volume measured, and its tracer concentration measured, to permit calculation of permeability.

All pulsatile, steady and transmural flow experiments were carried out for approximately 3-5 days prior to the measurement of transport.

*2.11 Data Analysis*

The statistical significance of differences was calculated using Student's t-test or ANOVA within each group of experiments and differences were deemed significant for $P<0.05$. There was no significant difference between bioreactors within each experimental series, so all bioreactors were treated as equal: pooled experimental data were compared with pooled control data from the same series.

# 3. Results

*3.1 Assessment of Endothelial Purity and Confluence*

The fraction of cells taking up DiI-acetylated-LDL was 99.72% (SEM 0.07%; n=5 isolations).

Glucose concentration in medium is shown as a function of time after seeding in Figure 2. Cells were exposed to CSS. As a control, concentrations were also measured in cell-free bioreactors.

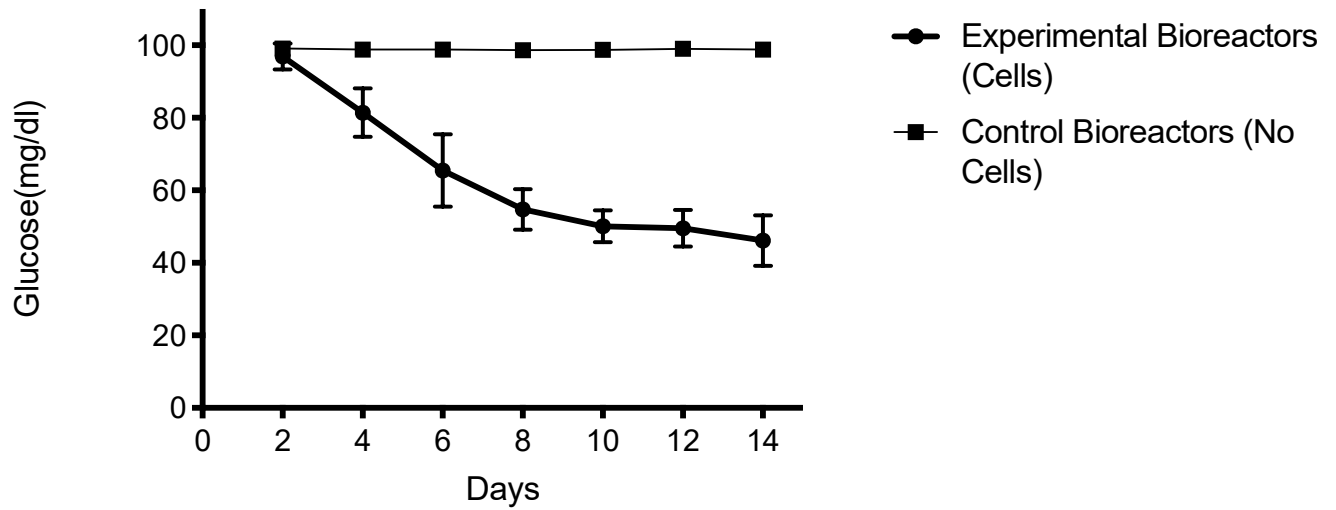

**Figure 2. Glucose consumption as a function of time after seeding**

Glucose concentration in the medium was measured every second day. Mean±SEM for all bioreactors used in subsequent experiments (n=27 for each condition).

We interpret the data as follows. PAECs reached confluence within 2-3 days after seeding. This led to a steady consumption of glucose for approximately the next 4 days, after which glucose consumption reduced and then reached a new but lower steady level, as expected for quiescent endothelium, at approximately 10 days.

Medium was changed and experimental measurements were started when glucose was depleted by approximately 50%, which took between 10 and 16 days in different bioreactors.

Figure 3 shows permeability to rhodamine-labelled albumin as a function of time after seeding in four bioreactors. The first measurement was made following cell seeding and all subsequent readings were taken after glucose concentration had reduced by 50%. (The time required for this varied between bioreactors, as noted above.) Cells were exposed to CSS. Permeability was also measured at equivalent times in cell-free bioreactors.

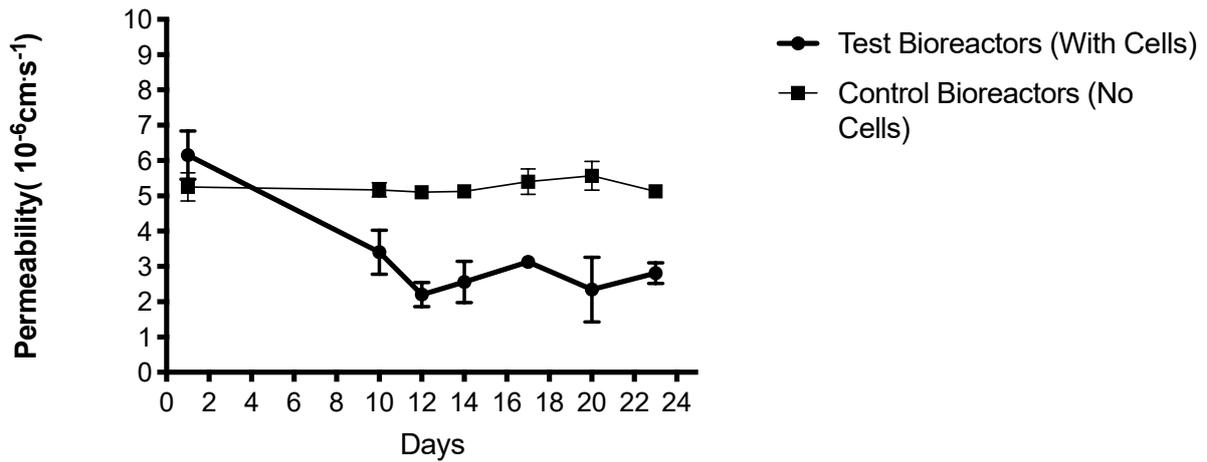

**Figure 3. Permeability to rhodamine-labelled albumin as a function of time after seeding**

Measurements were made in four bioreactors with cells and four without cells. Mean±SEM.

Permeability in experimental and control bioreactors was essentially identical at day 1. Subsequent measurements in seeded bioreactors averaged less than half of those in control bioreactors and their level was approximately constant from 10 days onwards. This is consistent with a tight barrier having been formed by the time that glucose levels had depleted by 50%.

Figure 4 shows SEM images of endothelial monolayers, viewed *en face* on a single fibre at the end of a series of experiments. Contiguous endothelial cells and boundaries are evident and the distinctive appearance of the unseeded fibre is not visible at any location or magnification, demonstrating that repeated permeability measurements and changes in flow do not substantially disrupt the confluent monolayer established soon after seeding.

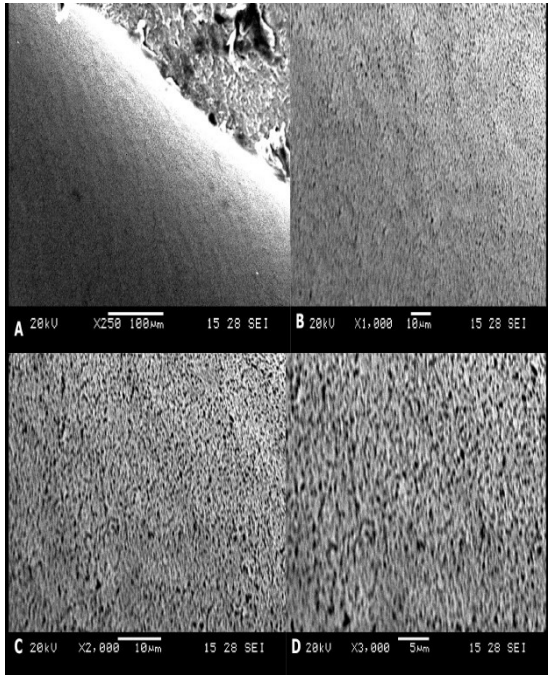

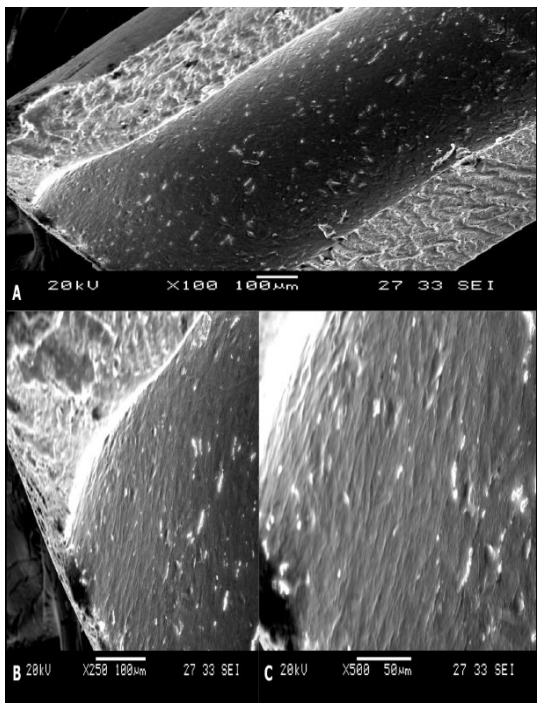

**Figure 4. Scanning electron micrographs of a confluent monolayer on a hollow fibre**

SEM images taken at increasing magnification show the luminal surface of fibres without cells (top panel, A-D) and with cells after prolonged experiments (bottom panel A-C).

*3.2 Replication of Previous Influences on Permeability*

3.2.1 Thrombin and L-NAME

Thrombin significantly increased permeability to 8.26 ± 0.75 x10-06 cm/s (n = 9) compared to chronic shear stress without thrombin (3.27 ± 0.425 x10-06 cm/s; n = 10; p < 0.0001) (Figure 5).

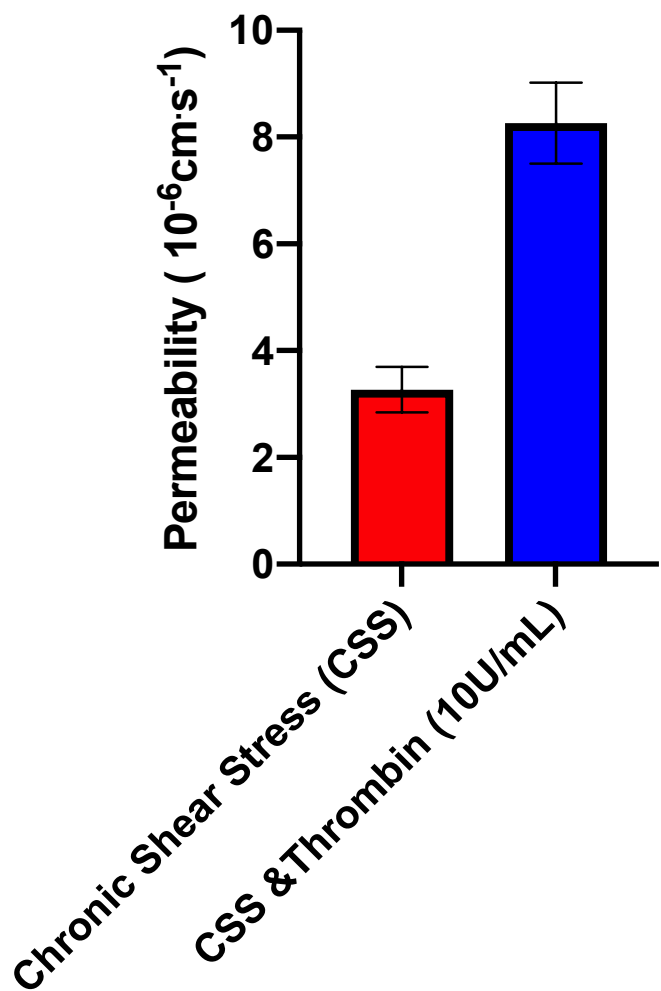

**Figure 5. The effect of thrombin on endothelial permeability under chronic pulsatile shear**

Permeability after thrombin (10 U/mL) was allowed to circulate for 1 h under CSS, and under CSS without thrombin. Thrombin significantly increased permeability to 8.26 ± 0.75 x$10^{-06}$ cm/s (n = 9, p < 0.0001) compared to the chronic shear stress alone (3.27 ± 0.425 x$10^{-06}$ cm/s; n = 10). Mean±SEM.

Permeability with L-NAME (5.62±0.62 x 10-06 cm/s; n = 9) was also significantly higher than permeability for controls (2.89 ± 0.18 x 10-06 cm/s; n = 9; p < 0.0001) (Figure 6).

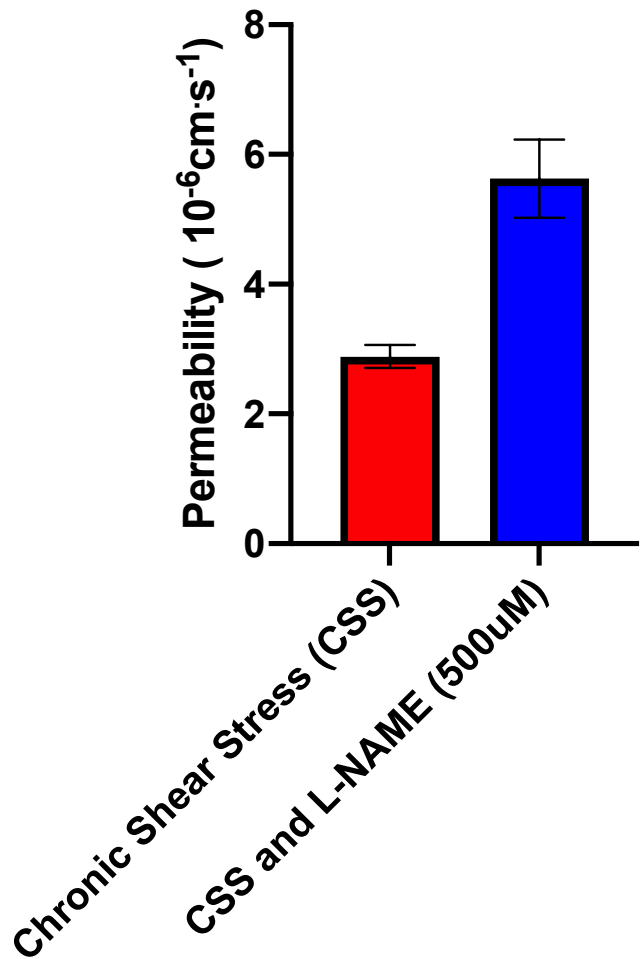

**Figure 6. The effect of nitric oxide inhibition on endothelial permeability under chronic pulsatile shear**

Permeability after L-NAME (500 µM) was allowed to circulate for 24 h under CSS, and under CSS without L-NAME. There was a statistically significant increase in permeability ($p < 0.0001$) with L-NAME ($5.62 \pm 0.62 \times 10^{-06}$ cm/s; n = 9) compared to chronic shear stress conditions without L-NAME ($2.89 \pm 0.18 \times 10^{-06}$ cm/s; n = 9). Mean±SEM.

3.2.2 Acute versus chronic shear stress

Permeability was increased by acute pulsatile shear ($8.78 \pm 0.62 \times 10^{-06}$ cm/s; n = 14; $p < 0.0001$) (Figure 7) and decreased by chronic pulsatile shear ($2.57 \pm 0.24 \times 10^{-06}$ cm/s; n = 9; $p < 0.0001$) (Figure 8) compared to static conditions ($5.09 \pm 0.25 \times 10^{-06}$ cm/s$^{-1}$; n = 13).

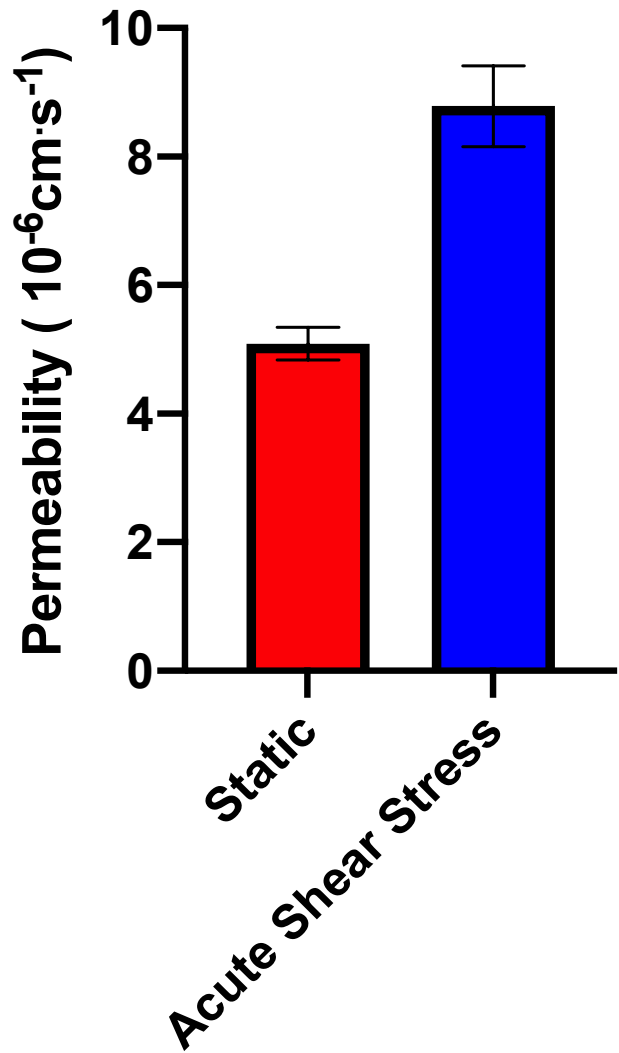

**Figure 1. The effect of acute pulsatile shear stress on endothelial permeability in the bioreactor**

4 h of pulsatile flow (ASS) led to an increase in permeability (8.78 ± 0.62 x$10^{-06}$ cm/s; n = 14; p < 0.0001) compared to the permeability under static conditions (5.09 ± 0.25 x $10^{-06}$ cm/s; n = 13). Mean±SEM.

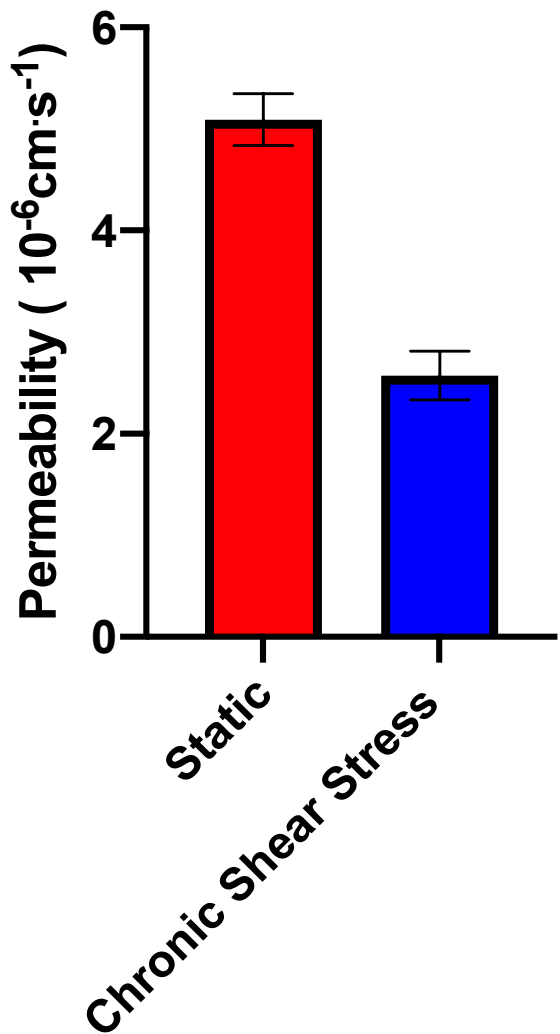

**Figure 8. The effect of chronic pulsatile shear stress on endothelial permeability in the bioreactor**

3-5 days of pulsatile flow (CSS) led to a decrease in permeability (2.57 ± 0.24 x$10^{-06}$ cm/s; n = 9; p < 0.0001) compared to static conditions (5.09 ± 0.25 x$10^{-06}$ cm/s; n = 13). Mean±SEM.

*3.3 Permeability with increased viscosity*

In bioreactors perfused with medium containing xanthan gum, permeability was not significantly increased by ASS compared to static conditions, although there was a trend in that direction (p = 0.0916), whilst CSS produced a significant decrease (p < 0.0001) (Figure 9, blue bars).

Figure 9 also shows data from the bioreactors perfused with conventional medium plus supplements and exposed to static conditions, ASS and CSS (red bars). There was no effect of increased viscosity under static conditions (p = 0.1718), whereas the presence of xanthan gum appeared to lower permeability under both ASS and CSS (p = 0.0067 and p = 0.0002, respectively).

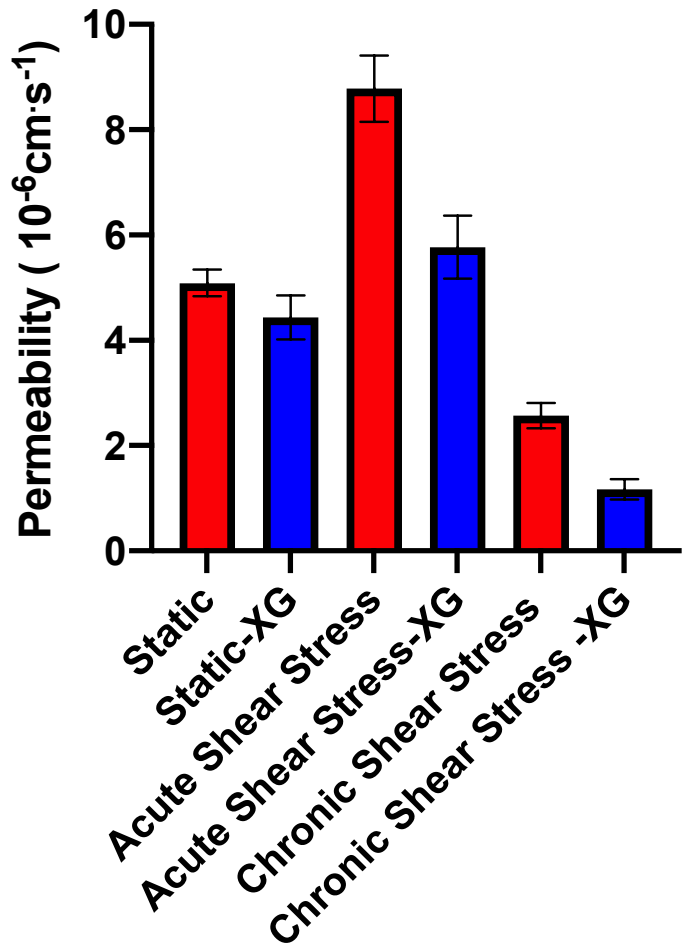

**Figure 9. The effect of acute pulsatile shear stress on endothelial permeability under increased viscosity in the hollow fibre bioreactor**

4 h of pulsatile flow (ASS) with xanthan gum (XG) led to a non-significant increase in permeability (5.77 ± 0.60 x $10^{-06}$ cm/s; p = 0.0933) and 3-5 days of pulsatile flow (CSS) with XG led to a significant decrease in permeability (1.17 ± 0.19 x 10-06 cm/s; n = 11; p < 0.0001) compared to static conditions (4.43 ± 0.41 x10-06 cm/s; n=7). (Blue bars). Earlier data obtained under the same conditions but without XG are shown for comparison. (Red bars). Permeability was higher under ASS without XG (n=14) than with it (n=7; p = 0.0067). The same trend was observed under CSS: permeability values were higher without XG (n = 9) than with it (n=11; p = 0.0002). Under static conditions, there was no significant difference with (n = 7) and without XG (n = 13; p = 0.1718). Mean±SEM.

*3.4 Permeability in Modified Bioreactors*

3.4.1 Effect of Flow Waveform

Figure 10 shows flow measured with zero, two and four 50 mL air-filled syringes attached to the bioreactor. Flow without any damping chambers resembled physiological flow in the aorta: there was forward-going flow over approximately one third of the cycle, then a period of reverse flow having smaller amplitude, and finally a period of low amplitude forward-going flow. The amplitude of the waveform decreased from 2 ml/s without damping to 0.1 ml/s with four syringes; due to the use of a positive displacement pump, the mean flow will have remained constant. The fundamental frequency was 0.55 Hz. For simplicity, the damped flow is termed "steady" here, although it still has an oscillatory component.

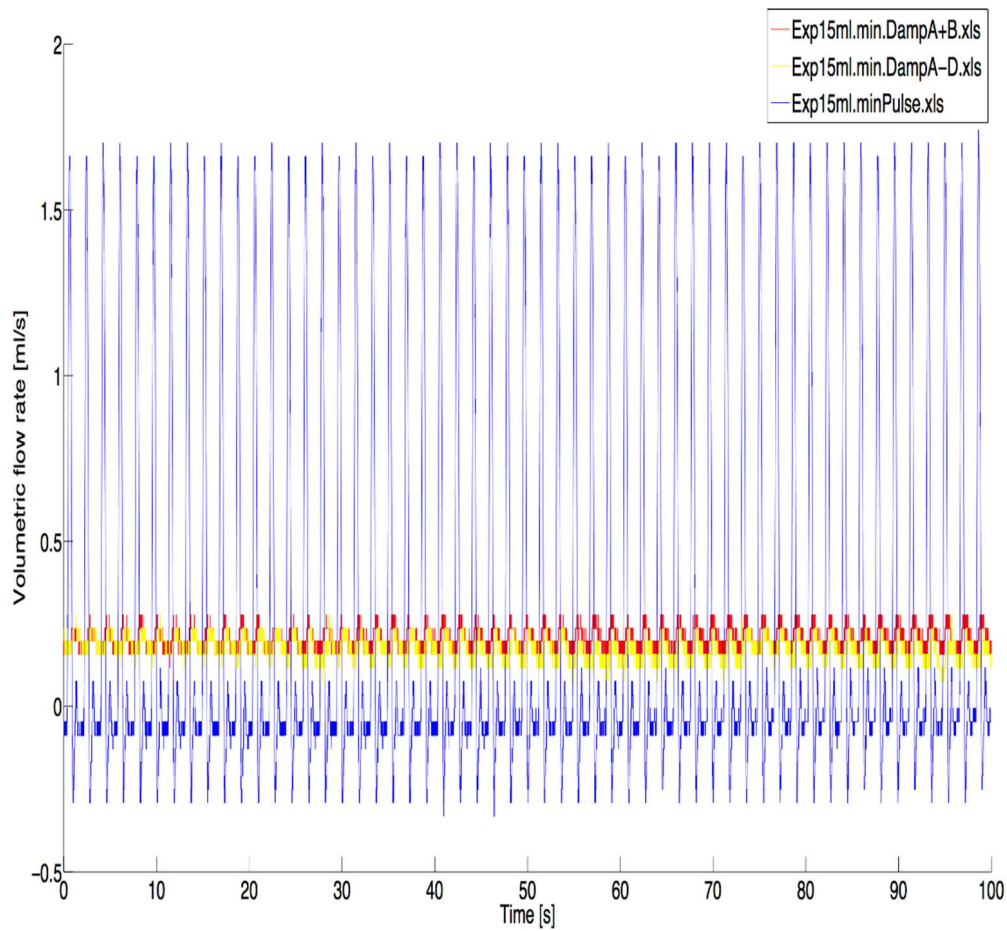

## Figure 10. Control and damped flow waveforms

Flow waveform, determined with a transit-time ultrasound flow probe, under normal perfusion (blue line) or with two (red line) or four air-filled syringes (yellow line) to provide damping.

There was a statistically significant increase in permeability when cells were exposed to chronic steady flow compared to the permeability values under chronic pulsatile flow (CSS; Figure 11). Permeability increased to $6.94 \pm 0.87 \times 10^{-06}$ cm/s (n = 15) under steady flow conditions, compared to pulsatile flow values of $3.15 \pm 0.218 \times 10^{-06}$ cm/s (n = 24; p < 0.0001).

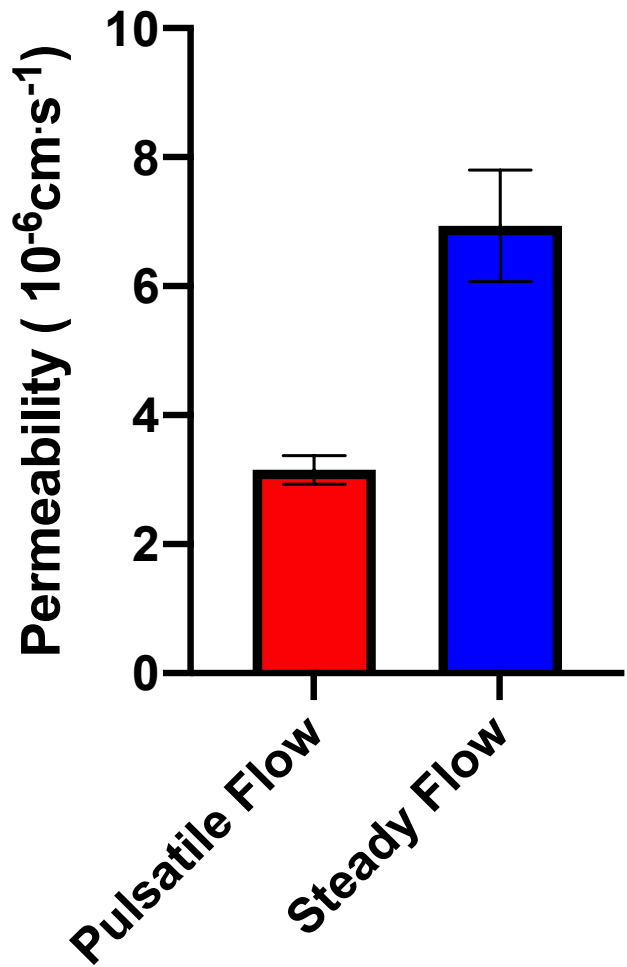

**Figure 11. The effect of flow waveform on endothelial permeability in the hollow fibre bioreactor**

Chronic steady flow led to a statistically significant increase in permeability (6.94 ± 0.87 x$10^{-06}$ cm/s; n = 15), compared to the normal chronic pulsatile flow (CSS; 3.15 ± 0.218 x$10^{-06}$ cm/s ; n = 24; p < 0.0001). Mean±SEM.

3.4.2 Effect of Transmural Flow

Flow from the intracapillary to the extracapillary space averaged approximately 0.1 ml/h, which will have had a negligible effect on the chronic pulsatile flow through the capillary lumen. (The transmural flow was driven by the same pulsatile pressure gradient as the luminal flow and would there have had an oscillatory component.)

Chronic pulsatile flow with the addition of transmural flow significantly increased permeability (8.56 ± 1.65 x $10^{-6}$ cm/s; n=15) compared to chronic pulsatile flow without transmural flow (2.62 ± 0.38 x $10^{-6}$ cm/s; n = 8; p = 0.0175) (Figure 12). (Strictly, these permeabilities should be termed "effective permeabilities" as they have a convective component.)

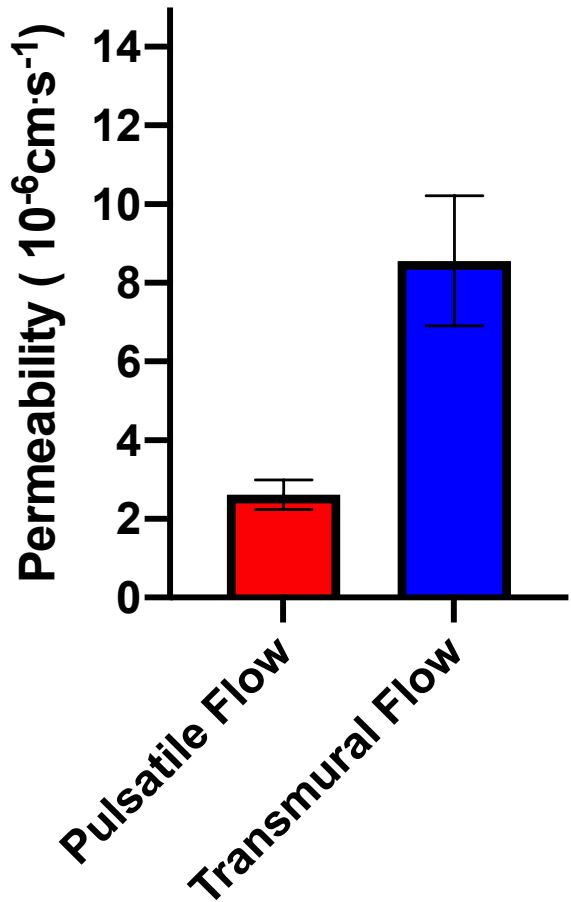

**Figure 12. The effect of transmural flow on endothelial effective permeability**

The introduction of transmural flow caused a significant increase in the effective permeability of the endothelium to rhodamine-labelled albumin (to 8.56 ± 1.65 x $10^{-6}$ cm/s; n = 15) when compared to pulsatile axial flow alone (2.62 ± 0.38 x $10^{-6}$ cm/s; n = 8; p = 0.0175). Mean±SEM.

A flow of 0.1 ml/h across the capillary surface area of 70 $cm^2$ is equivalent to a convective mass transfer coefficient of 0.4 x $10^{-6}$ cm/s, which is approximately 7% of the difference between these two results, so it could not directly account for the increase in albumin transport even if the reflection coefficient for albumin were zero. In practice, such coefficients are around 0.7 in cultured endothelial monolayers subject to an applied pressure difference, even in the absence of axial flow [28]. Hence we conclude that the increase in albumin transport is instead due to an increase in effective permeability of the monolayer induced by chronic transmural flow.

## 4. Discussion

The present study examined mechanical influences on the permeability of vascular endothelium *in vitro*. Arterial endothelial cells were used because the relation between fluids stresses and permeability in such vessels is relevant to the patchy anatomical distribution of atherosclerosis. They were taken from pigs rather than a smaller species because there is allometric scaling of WSS by body weight [29,30]; WSS normally experienced by pig arterial endothelium should be similar to that seen by human cells. Permeability was assessed by measuring the transport of albumin, a macromolecule that is widely used in such studies because it is relatively inert and present in plasma at high concentration. Receptor-mediated vesicular transport of albumin has been observed [31] but passive transport through normal or leaky intercellular junctions is thought to predominate for arterial endothelium [32]. Transport of the larger lipoproteins thought to be relevant to atherogenesis may [32] or may not [11, 33] dominantly occur by the same route. The cells were confluent and in a quiescent state before interventions commenced, as judged by measurements of permeability and glucose consumption.

*4.1 Replication of earlier effects of biological and mechanical stimuli*

The performance of the system, including the confluence of the monolayers, was checked by examining the effects of adding thrombin or L-NAME, which are known to increase permeability [34-36], and by applying acute or chronic shear stress [ASS or CSS], which respectively increase and decrease permeability [37], in other *in vitro* systems. If there had been frequent or large gaps between cells of the monolayer, such behaviour would not be detectable: the gaps are expected to dominate transport even if only 1 cell in 400 is missing [38].

Thrombin and L-NAME both altered permeability in the expected direction under chronic shear and the effects were large: permeability doubled with L-NAME and more than doubled with thrombin. Furthermore, ASS increased permeability to 1.7x that seen under static conditions whereas CSS halved it, values that are essentially identical to those obtained by Warboys et al [37] in the swirling well model. We can therefore at least say that the monolayer in the hollow fibre bioreactor is sufficiently confluent to test influences of other biological or mechanical agents, where effects are unknown, and can speculate that the monolayer in fact has very few defects.

*4.2 Effect of Viscosity*

Culture media used in most *in vitro* studies of endothelial permeability have a viscosity at 37 C of around 1 mPa.s. In contrast, the value for blood is around 4 mPa.s, depending on haematocrit. Furthermore, blood is a non-Newtonian, shear-thinning fluid whereas culture media are not. Obtaining physiological shear stresses at physiological shear rates therefore requires the addition of substances that increase viscosity, ideally in a non-Newtonian fashion, and a comparison of endothelial behaviour under flow with or without such additives can distinguish between responses to shear stress and shear rate.

Van den Broek et al [27] have shown that the food additive xanthan gum is ideal for the purpose. Medium containing only 0.69g/l of this bacterial polysaccharide approximated the viscosity and the shear-thinning behaviour of blood and was well tolerated by endothelial cells in prolonged culture. The additive was stable under prolonged flow in an *in vitro* system. The low mass concentration coupled with a high molecular weight (≈2 MDa) lead to only a small increment in molarity and osmotic pressure.

Since the present study used a positive displacement pump, the mean flow rate will not be affected by the change in viscosity. The flow is pulsatile and hence alteration of viscosity could theoretically alter the time-varying velocity profiles occurring within the capillaries. However, the Womersley number for our configuration is < 0.7 even without the xanthan gum; that is well within the viscous regime, so the flow is quasi-steady and hence velocity profiles (and therefore shear rates) should not be affected by the additive.

Xanthan gum had no influence on permeability under static conditions, consistent with a lack of direct effect on endothelial cells or on transport. (The large size of the molecule means that it will be significantly excluded from the intercellular junction and hence will likely not slow diffusion through it.) On the other hand, it significantly decreased permeability under both ASS and CSS, leading to a lack of significant difference between ASS and static conditions.

The increase in permeability caused by ASS and the decrease caused by CSS in the absence of xanthan gum thus both appear to depend on shear stress rather than shear rate: when shear stress and not shear rate was increased, both effects changed. The effect of CSS – which is to reduce permeability – became larger because WSS was increased. However, the normal effect of ASS, whch is to increase permeability, became smaller rather than larger. We speculate that this is because the initial increase and then decrease in permeability under shear is faster, as well as larger, when WSS is increased. Hence by 4 h, when transport under ASS was measured, permeability in the presence of the additive had dropped below the value seen without the additive.

Consistent with this speculation, Tarbell [10] reported the same phenomenon for albumin transport when flow was applied and then stopped 60 mins later. With both low and high applied shears, permeability increased when shear was applied and decreased when it ceased. At the higher shear, the initial increase in permeability was faster and larger than at the lower shear, but the return to baseline was also faster, leading to substantially (up to three-fold) *lower* permeability in the endothelium that had been exposed to higher shear, between 30 and 90 mins after the cessation of shear.

*4.3 Effects of Flow Waveform*

The pulsatility of the flow produced by the standard pump was damped by incorporating air chambers into the circuit. Four 50 mL syringes reduced flow rate amplitude 18-fold for the same mean flow. Chronic exposure of endothelium to this nearly steady flow resulted in higher permeability than exposure to chronic pulsatile flow. Note that steady flow is unusual in large arteries: the ratio of amplitude to the mean flow rate was <10% of that seen in the human aorta

[39], for example. Flow becomes steadier on progression through small arteries and arterioles, so it cannot be said that the pulsatile waveform is the more physiological in these smaller vessels.

Colgan et al [40] reached the opposite conclusion, that steady flow gives lower permeability than pulsatile flow. However, this inference was reached by assuming flow produced in the orbital shaker (swirling well) model is steady. Numerical and experimental studies [41-45] have shown that not to be true. Indeed, the flow may be more pulsatile than in the hollow fibre bioreactor and will certainly be less uniaxial. The authors also acknowledge that cells were replated between being flow exposure on the orbital shaker and the measurement of permeability, introducing an additional difference between the "steady" and pulsatile conditions. Adding damping chambers to the hollow fibre bioreactor provides a more controlled method.

If all endothelium behaves in the same way, then our data suggest that the endothelial barrier may be tighter in regions of pulsatile flow than in regions of steady flow (e.g. in arteries versus the microcirculation), and also that within arteries, areas experiencing more pulsatile flow might have locally tighter barriers.

*4.4 Transmural flow*

Fluid transport across the vessel wall occurs *in vivo* as a result of transmural gradients in oncotic and hydrostatic pressure; it advects solutes carried within plasma. The large majority of *in vitro* studies have not been able to include transmural flow for technical reasons, but it can be included when using the hollow fibre bioreactor simply by removing the constant-volume constraint on the extracapillary space, and hence lowering the hydrostatic pressure on the abluminal side of the capillary.

Previous studies by Tarbell et al. showed that applying [46] or increasing [47] the pressure difference across an endothelial monolayer produces at least three sequential changes in transmural flow: there is an initial, effectively instantaneous increase from the baseline value, followed by a reduction in flow to a lower value, still above baseline, over 30-120 minutes – the so-called "sealing effect" – and then a further increase in flow which plateaus after a period of approximately 5 h. Our measurements were made after applying a pressure difference for substantially longer than 5 h.

Transmural flow gives rise to a fluid dynamic shear stress on the endothelial cell membrane within intercellular junctions. In our system, the transmural flow velocity (Jv/A, where Jv is the volume flux) averaged $4 \times 10^{-7}$ cm/s. Tarbell et al. [47] obtained a ten-fold higher baseline value of $3-4 \times 10^{-6}$ cm/s, and used two simple models to show that the shear stress imposed on the walls of the cleft was of the order of 25-50 dynes/cm$^2$ [47]. In both models, shear was proportional to Jv, so we can assume values of approximately 2.5-5 dynes/cm$^2$ in our experiments, which is comparable in magnitude to the wall shear stress on the luminal surface of 3.75 dynes/cm$^2$. Tarbell et al. also estimated that the area of membrane in the clefts is 18% of the luminal membrane area. The substantial shear magnitude and large area over which it acts clearly could affect cell signalling. Transmural flow might thus alter monolayer properties through biological effects as well as by directly increasing solute transport through advection.

The addition of chronic transendothelial flow to chronic axial flow in the bioreactor increased transendothelial transport of rhodamine-labelled albumin more than could be explained by the addition of advection itself, even under the most favourable assumptions for the latter i.e. a reflection coefficient of zero. Neither was it due to stretch, because the capillary wall is rigid. We therefore conclude that transmural flow increased the permeability of the monolayer through a biological influence of junctional shear. Note that the effect is in the opposite direction to that of chronic luminal shear, which decreases permeability.

We did not examine the effect on albumin transport of short-term application of transmural flow but that has been investigated by DeMaio et al. [46], who measured albumin transport for 1 h before and 4 h after a pressure difference was applied across the monolayer. The sealing effect lasted approximately 2 h, during which period Jv fell by around 70%. The effective permeability to albumin increased 3-fold immediately after pressure was applied, but this increase was only around one third of the average transmural flow velocity (Jv/A) and it decreased approximately in step with the latter over the sealing period; these data do not provide strong evidence for a flow-driven, biologically mediated increase in permeability.

The discrepancy with our own data may reflect differences in the experimental conditions, such as the absence of flow parallel to the endothelial surface or the application of a steady rather than a pulsatile pressure difference in the experiments of Tarbell et al. However, the discrepancy is also consistent with a difference between acute and chronic effects of transmural flow, as with luminal flow; chronic effects presumably have greater physiological relevance.

## 5. Conclusion

We conclude by examining the advantages and disadvantages of the hollow fibre bioreactor for investigating endothelial permeability. (Note that broadly the same comments will apply to the study of epithelial permeability.)

*5.1 Advantages*

The commercially available device has a large number of advantages and these can be increased by minor modification. The main advantage is that sterile culture can be maintained for many weeks. That allows investigation of responses that are slow to develop. For example, the response to chronic shear (days) is different from the response to acute application of shear (hours), which is more frequently investigated. It also allows multiple measurements to be made on a single culture, even if the responses are slow, greatly increasing throughput. For example, in our comparison of permeability under ASS, CSS and static conditions, 36 experiments were conducted using just three bioreactors.

A second advantage is that the closed system with fluid recirculating through many narrow capillaries means that a large area of cells can be maintained chronically with a relatively low volume of medium. The surface area-to-volume ratio is comparable to, say, a 12-well plate with medium depth of 3mm. Hence costly bioactive substances or tracers can be added to the medium and products of endothelial metabolism can be detected in it.

A third advantage is that cells can be added to the extracapillary space, making it possible to examine effects of co-culture on endothelial permeability. It is well established that glial cells can tighten the barrier properties of monolayers and we have shown the same for dendritic cells, which are related to glial cells and are found in the aortic intima [48].

A fourth advantage is the ease of testing for active transport: tracer is allowed to circulate for sufficiently long that equilibrium should be reached if transport is purely passive, and its concentrations in the intracapillary and extracapillary space are then compared. A greater concentration in the extracapillary space indicates active transport from the luminal to the abluminal side of the monolayer. Evidence consistent with transport was obtained in preliminary experiments [49].

A final advantage, critical for the present study, is that the unmodified and modified bioreactor are n excellent platforms for examining effects of mechanical stresses on endothelial permeability. *First*, it is possible to have a static control because the fluid in the extracapillary space can be circulated to provide nutrients without being in direct contact with the endothelium; conventional parallel-plate flow chambers can only compare different kinds of flow. *Second*, different levels of shear can be imposed simply by altering pump speed, a method used by Westmuckett et al [22] in their study of tissue factor inhibitor release from cells. *Third*, the pulsatile flow produced by the off-the-shelf device produces a waveform resembling the one occurring in large arteries *in vivo*. *Fourth*, the flow waveform can be damped easily by adding air chambers between the pump and cartridge, allowing comparison of pulsatile and nearly steady flow. *Fifth*, the applied shear is the same at almost all locations: because of the small capillary diameter, Reynolds numbers are low and consequently entrance lengths are short, whilst the round cross section of the capillaries avoids the edge effects seen in parallel-plate flow chambers and swirled wells [42,45,50]. *Sixth*, altered viscosity can be used to distinguish effects of shear stress and shear rate because flow magnitude and waveform produced by the positive displacement pump should be unaffected and because the low Womersley number means that flow profiles should also be unaffected. *Seventh*, a minor modification allows a transmural flow to occur, in combination with luminal flow; hitherto, that has only been possible with complex, custom-built devices [51].

A further modification that could be considered is the introduction of compliant capillary walls, allowing effects of cyclic circumferential stretch to be examined. Preliminary studies of how this might be done are presented elsewhere [49].

*5.2 Disadvantages*

The first disadvantage – that albumin transport is much faster than *in vivo* – is common to all *in vitro* systems. It has been known for many years that endothelial permeability *in vivo* is on the order of $10^{-8}$ cm/s whereas *in vitro* it is on the order of $10^{-6}$ cm/s, as in the present study [28]. *In vitro*, cells of the monolayer are removed from their natural environment in terms of other cell types (including blood cells), the chemical milieu, the mechanical stresses to which the cells are exposed, and the extracellular matrix to which the cells adhere. Numerous attempts have been made to identify the factors responsible. It is likely that several are involved, but we and others have demonstrated that a substantial reduction in permeability to albumin can be achieved by adding sphingosine-1-phosphate to the perfusate [48,52]; this signalling lipid is normally carried by and released from circulating erythrocytes and platelets [53] and can account for the barrier tightening effect of platelet conditioned medium [54]. Orosomuciod, a circulating glycoprotein, is another candidate [55].

A second disadvantage is that the bioreactor is thought to produce flows that are aligned with the fibre axis and, by implication, with the long axis of the endothelial cells. (The diameter of the capillary is probably too large for the capillary itself to cause orientation of the cells [56].) The capillaries in the FiberCell® bioreactor actually have a wavy configuration but the radius of curvature is very long compared to the radius of the capillary [57]. That, coupled with the low Reynolds number, means the Dean number will also be low. In the absence of time-varying Dean vortices, flow direction is unlikely to change direction during each cycle. That is a limitation because of the recent focus on the role of multidirectional flow in atherogenesis; transverse WSS, a metric that captures components of the flow at right angles to the mean vector, correlates strongly with patterns of permeability and disease *in vivo* [58,59] and with permeability and inflammation *in vitro* [11,60,61].

Mandrycky et al [62] have recently published experimental and numerical evidence that secondary flows can be generated in even smaller capillaries than those used here by adding torsion. The theoretical basis for this has been debated [63,64]. The experimental data show a dependence of the secondary flow magnitude on axial flow velocity, which varies over the cycle in the bioreactor. Hence it may be possible to obtain multidirectional flow by incorporating torsion of the capillaries, for example by rotating the inlet and outlet manifolds relative to each other.

Finally, the major disadvantage is that the cells cannot be imaged before, during or between experiments. That is why monolayer confluence has to be assessed indirectly, through measurements of glucose consumption, permeability, or the response to biological and mechanical agents known to alter permeability. Responses of cells during individual experiments can be assessed if they are characterised by the release of chemicals into the medium. After the final experiment in each bioreactor, cells can be collected by trypsinisation and analysed [Ebrahim], or they can be imaged *in situ*, with capillaries being sliced as described above to expose the cells lining them. A complication when imaging the cells by fluorescence microscopy is the strong autofluorescence from the capillary wall which, unusually, is greater at longer wavelengths (> 400 nm). As a demonstration of feasibility, Figure 13 shows an image of the nuclei of endothelial cells lining a fibre, obtained with the UV-excited stain Hoechst 33342.

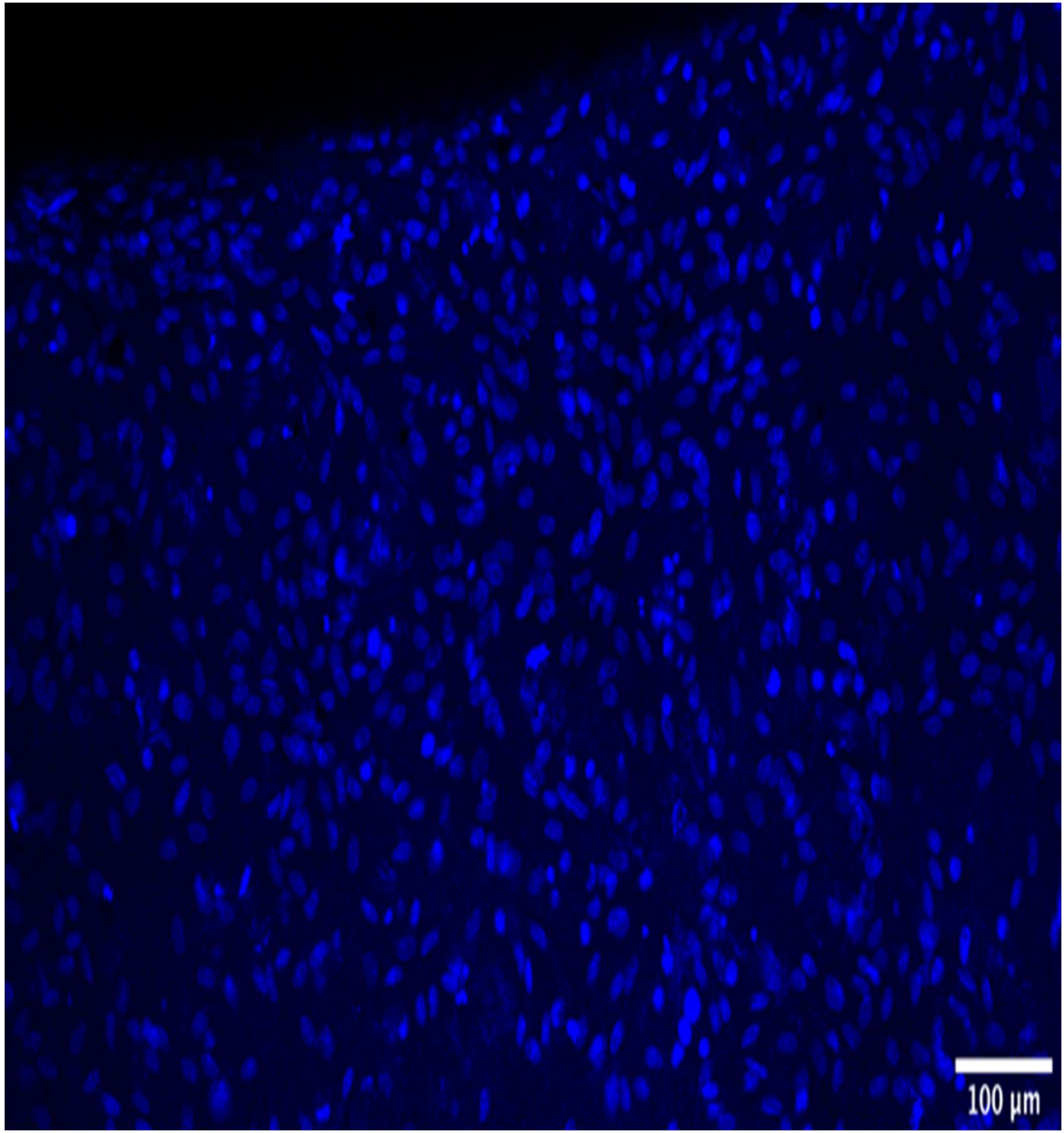

**Figure 13. Endothelial cell nuclei on a hollow fibre**

Cell nuclei on the luminal surface of a hollow fibre were stained for 10 mins with Hoescht 33342 nuclear stain (1:1000 dilution) and imaged using a Zeiss LSM-510 confocal microscope. The monolayer could be imaged without visible background autofluorescence due to the use of short wavelengths. The sample was from a bioreactor that experienced chronic shear stress for approximately 20 days; it was fixed in paraformaldehyde as described for the SEM samples.

## 6. Acknowledgements

We thank Y. Chooi, M. Ghim, A.A.E. Hunt and M. Ardakani for help with tracer preparation, cell isolation, flow measurement and electron microscopy, respectively. The work was funded by a BHF Centre of Research Excellence PhD studentship to SGG and a BHF Programme Grant to PDW.

References

1. Anichkov NN. Experimental arteriosclerosis in animals. In: Cowdry EV, editor, Arteriosclerosis: A survey of the problem. New York: MacMillan Publishing; 1933. p. 271–322.

2. Tarbell JM. Mass transport in arteries and the localization of atherosclerosis. Annu. Rev. Biomed. Eng. 2003;5:79–118.

3. Weinberg PD. Rate-limiting steps in the development of atherosclerosis: the response-to-influx theory. J Vasc Res. 2004;41:1-17.

4. Weinberg PD. Haemodynamic Wall Shear Stress, Endothelial Permeability and Atherosclerosis-A Triad of Controversy. Front Bioeng Biotechnol. 2022;10:836680.
H1543.

5. Staughton TJ, Lever MJ, Weinberg PD. Effect of altered flow on the pattern of permeability around rabbit aortic branches. Am J Physiol Heart Circ Physiol. 2001;281:H53-H59.

6. Nam D, Ni CW, Rezvan A, Suo J, Budzyn K, Llanos A, Harrison D, Giddens D, Jo H. Partial carotid ligation is a model of acutely induced disturbed flow, leading to rapid endothelial dysfunction and atherosclerosis. Am J Physiol Heart Circ Physiol. 2009;297:H1535-H1543.

7. Carew TE, Patel DJ. Effect of tensile and shear stress on intimal permeability of the left coronary artery in dogs. Atherosclerosis. 1973;18:179-189.

8. Lever MJ, Tarbell JM, Caro CG. The effect of luminal flow in rabbit carotid artery on transmural fluid transport. Exp Physiol. 1992 Jul;77(4):553-63.

9. Eaton BM, Toothill VJ, Davies HA, Pearson JD, Mann GE. Permeability of human venous endothelial cell monolayers perfused in microcarrier cultures: effects of flow rate, thrombin, and cytochalasin D. J Cell Physiol. 1991;149:88-99.

10. Jo H, Dull RO, Hollis TM, Tarbell JM. Endothelial albumin permeability is shear dependent, time dependent, and reversible. Am J Physiol. 1991;260:H1992-H1996.

11. Ghim M, Alpresa P, Yang SW, Braakman ST, Gray SG, Sherwin SJ, van Reeuwijk M, Weinberg PD. Visualization of three pathways for macromolecule transport across cultured endothelium and their modification by flow. Am J Physiol Heart Circ Physiol. 2017;313:H959-H973

12. Gao F, Sun H, Li X, He P. Leveraging avidin-biotin interaction to quantify permeability property of microvessels-on-a-chip networks. Am J Physiol Heart Circ Physiol. 2022;322:H71-H86

13. Dull RO, Jo H, Sill H, Hollis TM, Tarbell JM. The effect of varying albumin concentration and hydrostatic pressure on hydraulic conductivity and albumin permeability of cultured endothelial monolayers. Microvasc Res. 1991;41:390-407.

14. Knazek RA, Gullino PM, Kohler PO, Dedrick RL. Cell Culture on Artificial Capillaries: An Approach to Tissue Growth in vitro. Science 1972;178:65-67.

15. Eghbali H, Nava MM, Mohebbi-Kalhori D, Raimondi MT. Hollow fiber bioreactor technology for tissue engineering applications. Int J Artif Organs. 2016;39:1-15.

16. Stanness KA, Guatteo E, Janigro D. A dynamic model of the blood-brain barrier "in vitro". Neurotoxicology. 1996;17:481-96.

17. Terashima M, Fujita Y, Sugano K, Asano M, Kagiwada N, Sheng Y, Nakamura S, Hasegawa A, Kakuta T, Saito A. Evaluation of water and electrolyte transport of tubular epithelial cells under osmotic and hydraulic pressure for development of bioartificial tubules. Artif Organs. 2001;25:209-212.

18. Stanness KA, Westrum LE, Fornaciari E, Mascagni P, Nelson JA, Stenglein SG, Myers T, Janigro D. Morphological and functional characterization of an in vitro blood-brain barrier model. Brain Res. 1997;771:329-342

19. Neuhaus W, Lauer R, Oelzant S, Fringeli UP, Ecker GF, Noe CR. A novel flow based hollow-fiber blood-brain barrier in vitro model with immortalised cell line PBMEC/C1-2. J Biotechnol. 2006;125:127-141

20. Salvetti F, Cecchetti P, Janigro D, Lucacchini A, Benzi L, Martini C. Insulin permeability across an in vitro dynamic model of endothelium. Pharm Res. 2002;19:445-450.

21. Walsby E, Buggins A, Devereux S, Jones C, Pratt G, Brennan P, Fegan C, Pepper C. Development and characterization of a physiologically relevant model of lymphocyte migration in chronic lymphocytic leukemia. Blood. 2014;123:3607-3617.

22. Westmuckett AD, Lupu C, Roquefeuil S, Krausz T, Kakkar VV, Lupu F. Fluid flow induces upregulation of synthesis and release of tissue factor pathway inhibitor in vitro. Arterioscler Thromb Vasc Biol. 2000;20:2474-2482.

23. Milovanova T, Manevich Y, Haddad A, Chatterjee S, Moore JS, Fisher AB. Endothelial cell proliferation associated with abrupt reduction in shear stress is dependent on reactive oxygen species. Antioxid Redox Signal. 2004;6:245-258.

24. Bogle RG, Baydoun AR, Pearson JD, Mann GE. Regulation of L-arginine transport and nitric oxide release in superfused porcine aortic endothelial cells. J Physiol. 1996;490:229-241.

25. Du W, Ren L, Hamblin MH, Fan Y. Endothelial Cell Glucose Metabolism and Angiogenesis. Biomedicines. 2021;9:147.

26. Clarke LA, Zahra Mohri, Weinberg PD. High throughput en face mapping of arterial permeability using tile scanning confocal microscopy. Atherosclerosis. 2012;224:417-425.

27. van den Broek CN, Pullens RA, Frøbert O, Rutten MC, den Hartog WF, van de Vosse FN. Medium with blood-analog mechanical properties for cardiovascular tissue culturing. Biorheology. 2008;45:651-661.

28. Suttorp N, Hessz T, Seeger W, Wilke A, Koob R, Lutz F, Drenckhahn D. Bacterial exotoxins and endothelial permeability for water and albumin in vitro. Am J Physiol. 1988;255:C368-C376.

29. Greve JM, Les AS, Tang BT, Draney Blomme MT, Wilson NM, Dalman RL, Pelc NJ, Taylor CA. Allometric scaling of wall shear stress from mice to humans: quantification using cine phase-contrast MRI and computational fluid dynamics. Am J Physiol Heart Circ Physiol. 2006;291:H1700-1708.

30. Weinberg PD, Ross Ethier C. Twenty-fold difference in hemodynamic wall shear stress between murine and human aortas. J Biomech. 2007;40:1594-1598.

31. Schnitzer JE. gp60 is an albumin-binding glycoprotein expressed by continuous endothelium involved in albumin transcytosis. Am J Physiol. 1992;262:H246-H254.

32. Cancel LM, Fitting A, Tarbell JM. In vitro study of LDL transport under pressurized (convective) conditions. Am J Physiol Heart Circ Physiol. 2007;293:H126-H132.

33. Ramírez CM, Zhang X, Bandyopadhyay C, Rotllan N, Sugiyama MG, Aryal B, Liu X, He S, Kraehling JR, Ulrich V, Lin CS, Velazquez H, Lasunción MA, Li G, Suárez Y, Tellides G, Swirski FK, Lee WL, Schwartz MA, Sessa WC, Fernández-Hernando C. Caveolin-1 Regulates Atherogenesis by Attenuating Low-Density Lipoprotein Transcytosis and Vascular Inflammation Independently of Endothelial Nitric Oxide Synthase Activation. Circulation. 2019;140:225-239.

34. Malik AB, Lo SK, Bizios R. Thrombin-induced alterations in endothelial permeability. Ann N Y Acad Sci. 1986;485:293-309.

35. Kang H, Cancel LM, Tarbell JM. Effect of shear stress on water and LDL transport through cultured endothelial cell monolayers. Atherosclerosis. 2014;233:682-690

36. Ghim M, Mohamied Y, Weinberg PD. The Role of Tricellular Junctions in the Transport of Macromolecules Across Endothelium. Cardiovasc Eng Technol. 2021;12:101-113.

37. Warboys CM, Eric Berson R, Mann GE, Pearson JD, Weinberg PD. Acute and chronic exposure to shear stress have opposite effects on endothelial permeability to macromolecules. Am J Physiol Heart Circ Physiol. 2010;298:H1850-H1856.

38. Tzeghai G, Ganatos P, Pfeffer R, Weinbaum S, Nir A. A theoretical model to study the effect of convection and leaky junctions on macromolecule transport in artery walls. J Theor Biol. 1986;121:141-162.

39. Hope TA, Markl M, Wigström L, Alley MT, Miller DC, Herfkens RJ. Comparison of flow patterns in ascending aortic aneurysms and volunteers using four-dimensional magnetic resonance velocity mapping. J Magn Reson Imaging. 2007;26:1471-1479.

40. Colgan OC, Ferguson G, Collins NT, Murphy RP, Meade G, Cahill PA, Cummins PM. Regulation of bovine brain microvascular endothelial tight junction assembly and barrier function by laminar shear stress. Am J Physiol Heart Circ Physiol. 2007;292:H3190-H3197.

41. Berson RE, Purcell MR, Sharp MK. Computationally determined shear on cells grown in orbiting culture dishes. Adv Exp Med Biol. 2008;614:189–98.

42. Salek MM, Sattari P, Martinuzzi RJ. Analysis of Fluid Flow and Wall Shear Stress Patterns Inside Partially Filled Agitated Culture Well Plates. Ann Biomed Eng. 2012;40:707–28.

43. Alpresa P, Sherwin S, Weinberg P, van Reeuwijk M. Orbitally shaken shallow fluid layers. I. Regime classification. Phys Fluids. 2018;30:032107.

44. Warboys CM, Ghim M, Weinberg PD. Understanding mechanobiology in cultured endothelium: A review of the orbital shaker method. Atherosclerosis. 2019;285:170-177.

45. Arshad M, Rowland EM, Riemer K, Sherwin SJ, Weinberg PD. Improvement and validation of a computational model of flow in the swirling well cell culture model. Biotechnol Bioeng. 2022 Jan;119(1):72-88.

46. DeMaio L, Tarbell JM, Scaduto RC Jr, Gardner TW, Antonetti DA. A transmural pressure gradient induces mechanical and biological adaptive responses in endothelial cells. Am J Physiol Heart Circ Physiol. 2004;286:H731-H741.

47. Tarbell JM, Demaio L, Zaw MM. Effect of pressure on hydraulic conductivity of endothelial monolayers: role of endothelial cleft shear stress. J Appl Physiol (1985). 1999;87:261-268.

48. Warboys CM, Overby DR, Weinberg PD. Dendritic cells lower the permeability of endothelial monolayers. Cell Mol Bioeng. 2012;5:184-193.

49. Gray SG. Investigating biomechanical determinants of endothelial permeability in a hollow fibre bioreactor. PhD Thesis, Imperial College London 2021. https://spiral.imperial.ac.uk/handle/10044/1/100942

50. Viegas KD, Dol SS, Salek MM, Shepherd RD, Martinuzzi RM, Rinker KD. Methicillin resistant Staphylococcus aureus adhesion to human umbilical vein endothelial cells demonstrates wall shear stress dependent behaviour. Biomed Eng Online. 2011;10:20.

51. Sill HW, Chang YS, Artman JR, Frangos JA, Hollis TM, Tarbell JM. Shear stress increases hydraulic conductivity of cultured endothelial monolayers. Am J Physiol. 1995;268:H535-H543.

52. Curry FE, Clark JF, Adamson RH. Erythrocyte-derived sphingosine-1-phosphate stabilizes basal hydraulic conductivity and solute permeability in rat microvessels. Am J Physiol Heart Circ Physiol. 2012;303:H825-H834.

53. Warboys CM, Weinberg PD. S1P in the development of atherosclerosis: roles of hemodynamic wall shear stress and endothelial permeability. Tissue Barriers. 2021;9:1959243.

54. Schaphorst KL, Chiang E, Jacobs KN, Zaiman A, Natarajan V, Wigley F, Garcia JG. Role of sphingosine-1 phosphate in the enhancement of endothelial barrier integrity by platelet-released products. Am J Physiol Lung Cell Mol Physiol. 2003;285:L258-L267.

55. Haraldsson B, Rippe B. Orosomucoid as one of the serum components contributing to normal capillary permselectivity in rat skeletal muscle. Acta Physiol Scand. 1987;129:127-135.

56. Dessalles CA, Leclech C, Castagnino A, Barakat AI. Integration of substrate- and flow-derived stresses in endothelial cell mechanobiology. Commun Biol. 2021;4:764.

57. Figure 3b in [https://www.fibercellsystems.com/wp-content/uploads/2015/06/FiberCell-Systems-User%E2%80%99s-Manual-r2.pdf](https://www.fibercellsystems.com/wp-content/uploads/2015/06/FiberCell-Systems-User%E2%80%99s-Manual-r2.pdf)

58. Peiffer V, Sherwin SJ, Weinberg PD. Computation in the rabbit aorta of a new metric - the transverse wall shear stress - to quantify the multidirectional character of disturbed blood flow. J Biomech. 2013;46:2651-2658.

59. Mohamied Y, Rowland EM, Bailey EL, Sherwin SJ, Schwartz MA, Weinberg PD. Change of direction in the biomechanics of atherosclerosis. Ann Biomed Eng. 2015;43:16-25.

60. Ghim M, Pang KT, Arshad M, Wang X, Weinberg PD. A novel method for segmenting growth of cells in sheared endothelial culture reveals the secretion of an anti-inflammatory mediator. J Biol Eng. 2018;12:15.

61. Ghim M, Pang KT, Burnap SA, Baig F, Yin X, Arshad M, Mayr M, Weinberg PD. Endothelial cells exposed to atheroprotective flow secrete follistatin-like 1 protein which reduces transcytosis and inflammation. Atherosclerosis. 2021;333:56-66.

62. Mandrycky C, Hadland B, Zheng Y. 3D curvature-instructed endothelial flow response and tissue vascularization. Sci Adv. 2020;6:eabb3629.

63. Wang C. On the low-Reynolds-number flow in a helical pipe. J Fluid Mech 1981;108:185-194.

64. Germano M. On the effect of torsion on a helical pipe flow. Journal of Fluid Mechanics 1982;125:1-8.

65. Ebrahim AS, Carion TW, Strand E, Young LA, Shi H, Berger EA. Application of a Flow-Based Hollow-Fiber Co-Culture System to Study Cellular Influences under Hyperglycemic Conditions. Sci Rep. 2019;9:3771.